\documentclass[12pt]{article}

\usepackage{graphicx}
\usepackage{amssymb}
\usepackage{amsthm}
\usepackage{booktabs}
\usepackage{rotating}
\usepackage{multirow}
\usepackage{eurosym}
\usepackage{amsfonts}
\usepackage{amsmath}
\usepackage{umoline}
\usepackage{caption}
\usepackage{subcaption}
\usepackage{color}
\usepackage{url}
\usepackage{authblk}
\usepackage[english]{babel}

\usepackage{pbox}
\usepackage{footnote}
\usepackage{multirow}
\usepackage{changepage}
\usepackage{tablefootnote}

\usepackage[margin = 2cm]{geometry}

\providecommand{\keywords}[1]{\textbf{\textit{Keywords:}} #1}

\usepackage{hyperref}

\usepackage[natbibapa]{apacite}

\usepackage{rotating}
\usepackage{lscape}
\usepackage{float}
\restylefloat{table} 
\usepackage{booktabs}
\usepackage{textcomp}
\usepackage{siunitx}
\usepackage{multirow}
\usepackage{makecell}
\usepackage{longtable}
 
\begin{document}

\title{Evolution of Credit Risk Using a Personalized Pagerank Algorithm for Multilayer Networks \footnote{\scriptsize NOTICE: This paper has been accepted for publication at the KDD MLF 2020 Workshop. Please cite this paper as follows: Cristi\'{a}n Bravo and Mar\'{i}a \'{O}skarsd\'{o}ttir. 2020. Evolution of Credit Risk Using a Personalized Pagerank Algorithm for Multilayer Networks. In Proceedings of Third KDD Workshop on Machine Learning in Finance, joint with 26th ACM SIGKDD Conference on Knowledge Discovery in Databases (KDD MLF 2020). ACM, New York, NY, USA, 8 pages. Changes resulting from the publishing process, such as peer review, editing, corrections, structural formatting, and other quality control mechanisms may not be reflected in this version of the document. Changes may have been made to this work since it was submitted for publication.}}

\author[1]{Cristi\'{a}n Bravo \thanks{Corresponding author: cbravoro@uwo.ca}}
\author[2]{Mar\'{i}a \'{O}skarsd\'{o}ttir}

\affil[1]{Department of Statistical and Actuarial Sciences, The University of Western Ontario, Western Science Centre, 1151 Richmond Street, London, ON, Canada.}

\affil[2]{Department of Computer Science, Reykjavik University, Reykjavík, Iceland.}

\date{}

\maketitle

\begin{abstract}
In this paper we present a novel algorithm to study the evolution of credit risk across complex multilayer networks. Pagerank-like algorithms allow for the propagation of an influence variable across single networks, and allow quantifying the risk single entities (nodes) are subject to given the connection they have to other nodes in the network. Multilayer networks, on the other hand, are networks where subset of nodes can be associated to a unique set (layer), and where edges connect elements either intra or inter networks. Our personalized PageRank algorithm for multilayer networks allows for quantifying how credit risk evolves across time and propagates through these networks. By using bipartite networks in each layer, we can quantify the risk of various components, not only the loans. We test our method in an agricultural lending dataset, and our results show how default risk is a challenging phenomenon that propagates and evolves through the network across time.
\end{abstract}


\section{Introduction}

Within credit risk measurement, the problem of risk propagation has been a tough open question. Original models for credit risk, such as the underlying Basel models which underpins all modern banking regulation \citep{BCBSExplanatory2005}, assume independence between borrowers and no risk correlation except for the one each product line (consumer loans, revolving, mortgages) inherently has with the economy as a whole. This is a very general, and imprecise, assumption which has been recognised as a standing issue for many years \citep{Thomas_2005}. Several studies exist which focus on measuring the impact of default correlation. Evidence of its impact has been shown in consumer lending \citep{fenech2015} and corporate lending \citep{duan2016}, but only recently have some studies focused on incorporating these relations explicitly when modelling credit risk, particularly in consumer credit risk \citep{oskarsdottir2019} and for small businesses \citep{calabrese2019}.

Most of these modern studies are based on applying some form of network analysis to lending datasets. Social Network Analytics in particular allows to explicitly and individually measure how risk propagates across a network, using algorithms such as Personalised PageRank \citep{page1999pagerank}. However, and motivating this work, it is natural to encounter problems where there is more than one single network connecting the members of the network. These complex networks are called \emph{Multilayer Networks} and will be the focus of this study.

 Multilayer networks arise naturally in many applications. Two nodes in a network can be connected by more than one type of edge. For example, two people can be connected because they are family and because they work at the same institution. Similarly, two people can be connected in two different social media platforms, such as Facebook and Twitter, and additionally, people who have an account on Facebook may or may not have a Twitter account. This means there may not be the same set of nodes in all networks, which further enhances the complexity of the problem. Within credit risk, multilayer connections have been observed between banks \citep{thurner2013debtrank} and shown to be relevant to determine the systemic risk of an economy, and worst yet, if multilayer effects are ignored, risk is ``drastically underestimated'' \citep{poledna2015multi}.

In this work, we focus our attention to multilayer networks in retail lending and present a novel personalized PageRank algorithm for multilayer networks. We build a complex and information-rich network using multilayer network theory and bipartite graphs on data that, in a single layer representation, would result in a vast number of cliques that are not very informative. We do this by using characteristics as the aggregator nodes, which allow for rich and complex network representations to arise whenever there are classes that can be leveraged to create connections. We then apply the personalized PageRank algorithm to the networks, using defaulted loans as the information source, in order to rank the nodes with respect to their exposure to default risk. This allows us to quantify the nodes' risk influence across this complex connected structure.

Instead of looking at the loans themselves, we focus on the nodes which loans are connected to. To motivate the methodology, we show the model performance over a set of agricultural loans. In this context, two networks arise naturally: On the one hand, the product the agribusiness produces creates correlated risks as pests or other product-specific phenomena can lead to effects across all producers of the same product. On the other hand, the district in which the agribusiness operates also leads to correlated risks, as local weather events can impact production output and prices. Our research questions thus are:

\begin{enumerate}
    \item What is the best way to represent lending-related multilayer networks in order to apply propagation algorithms?
    \item How does default risk propagate across this network?
    \item How does this correlated default risk evolve across time and how does it compare to spot default rates?
\end{enumerate}

In what follows we investigate the changes in risk exposure over ten years in order to quantify the impact of each node in a multilayer environment induced by connecting variables, and measure its evolution through time. 

The structure of this paper is the following: the next section positions this paper in current literature, both in retail lending and in network science, followed by the theoretical framework for multilayer networks in Section \ref{sec:MLN}. The following section proposes the Multiplex Personalized Pagerank Algorithm which will be the basis of this work. Section \ref{sec:Experiments} presents the dataset and the experimental results. The final section shows the conclusion of the paper.

\section{Previous works}\label{sec:Literature}

\subsection{Retail Credit Risk Measurement}
 
 Credit scoring is, without a doubt, one of the first applications of predictive analytics. The first statistical experiences come from 1941 \citep{anderson2019}, but it is not until the late 80s when it starts to become the ubiquitous tool it now is \citep{thomas2017}. Many papers exist dealing with how to measure credit risk, from simple applications to sophisticated machine learning models \citep{lessmann2015}. However very few of these works find their way into practical applications, mostly because stringent banking regulations require transparency and explainability which machine learning models cannot provide \citep{BCBSExplanatory2005}. Thus, most modern models are based on simple logistic regressions, which are then rescaled to be more palatable to users \citep{Siddiqi2017}. 
 
 One area where research has been able to contribute has been on the use of so-called Alternative Data \citep{cheney2008}, or the use of non-traditional, usually non-payment, data sources to enrich credit risk modelling. Examples of alternative data are utility payments \citep{Bravo_2013}, social media information \citep{zhang2016}, psychometrics \citep{arraiz2017}, and - more to the point of this work - network information \citep{oskarsdottir2019}.   
 
 This last source of data has been recently proved effective in the consumer retail sphere. Our previous work showed how effective network data coming from telecommunications information can be used to predict credit default, particularly in consumers with no previous credit history. Telco data has been extended into mainstream credit risk models as well: Experian, one of the three largest credit bureaus worldwide recently launched a tool \citep{renton2019} claiming to use ``positive telecom and utility payment history'' directly into their scores. Of course, the use of network data comes associated to new concerns, such as ethical issues on such data use and the legal agreements necessary for sharing such data. Other experiences using network data for credit risk modelling include scores based exclusively on telco data \citep{sanpedro2015}, and based exclusively on spatial dependence \citep{zhou2017}.
 
 These past experiences all focus on one source of relational data, but this is not a hard constraint. Multiple sources of data can be leveraged to create connections, usually through explicit networks, but also through implied (``ego'') networks \citep{mcauley2014}. Given the natural and ubiquitous occurrence of networks in data, a rigorous procedure to extract information from these multi-source (\emph{multilayer}) networks becomes necessary.
 
\subsection{Multilayered networks}
Many systems that arise naturally in e.g., biology, sociology, finance and economics, consist of entities that interact with each other, and can therefore be represented by networks. Network science has evolved as an essential tool to describe and analyse such systems \cite{barabasi2016network}.

The interactions between entities can be complicated, with various types of connections between entities in different subsystems, for which multilayer network theory has been developed in the last few years \cite{kivela2014multilayer}. In addition to nodes and edges that make up regular single layer networks, multilayer networks have various layers. In the most general case, any node can belong to any subset of layers, edges can connect any two nodes in any layer, or pairs of nodes between any two layers.

Different forms of multilayer networks exist, depending on the characteristics of nodes and edges within and between the various layers and the constraints which they are subjected to. For example, in multiplex networks the node sets are the same in each layer and edges between layers connect the same node in two different layers.

A central theme in network science is centrality, that is, to identify the nodes that are most important or influential, in some regard. The most common centrality measures are degree, closeness, betweenness and eigenvector --or PageRank-- centrality. Degree counts the number of edges incident to a node and is both conceptually and computationally the simplest. Closeness measures the average distance of a node to all other nodes in the network and thus how close it is to the other nodes. Betweenness represents how often a node lies on the shortest path between the other nodes in the network.

PageRank centrality measures the influence a node has in a network. It depends on the number of nodes that link to a node and their PageRank centrality. It was originally designed to rank webpages in search engines but has been utilized in numerous applications \cite{kwak2010twitter,min2018behavior,lohmann2010eigenvector,botelho2011combining}. It is based on the stationary distribution of a random walker in a network with a random teleportation through a restart vector. The restart vector of the equation can be manipulated to steer the random walk towards a set of source nodes. This is known as Personalized PageRank \citep{page1999pagerank}. It has been used in various applications to rank nodes with respect to nodes of influence so that the ones that are closer to the source obtain a higher score.

PageRank centrality has been generalised to multilayer and, in particular, multiplex networks as follows.
\citet{iacovacci2016functional,iacovacci2016extracting} proposed functional Multiplex PageRank which is capable of capturing non-linear effects caused by different types of links between nodes.  
\citet{halu2013multiplex} extended the PageRank centrality measure to multiplex network by computing a traditional PageRank in one layer of the network and using the resulting ranking as input for the ranking on the second layer. Thus the first layer can contribute to the importance in the other layer.
\citet{pedroche2016biplex} generalised the notation of the classical PageRank in the style of multiplex networks and then extended them to multiplex networks.

Finally, a substantial body of research is devoted to the development of MuxViz, a tool to analyse and visualise multilayer networks \citep{de2015muxviz}.
In the development, several features of networks were extended to multilyaer networks, including the mathematical formulation, diffusion and centrality \citep{kivela2014multilayer, de2013mathematical, domenico2013centrality, gomez2013diffusion}. 
In these works, the authors show that the multiplex PageRank can be obtained in a similar way as the regular PageRank, using eigentensors. 
\begin{figure*}
    \centering
    \includegraphics[width=\linewidth]{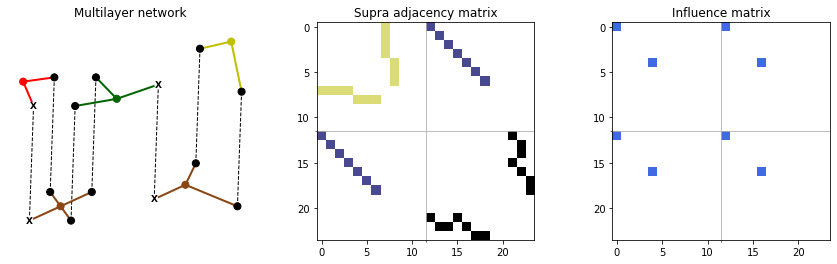}
    \caption{A multilayer network (left), its supra adjacency matrix (middle) and an influence matrix (right). The multilayer network has two layers which both contain a bipartite network.  There are seven common nodes that are in both layers (black). The common nodes in the first layer are connected with brown edges to brown nodes and the common nodes in the second layer are connected with red, green and yellow edges to red, green and yellow nodes, respectively. The inter edges (dashed) connect the common nodes to themselves in the other layer. In the supra adjacency matrix the upper left submatrix (yellow) denotes the adjacency matrix for the first layer, the lower right matrix (black) is the adjacency matrix for the second layer. The upper right and lower left submatrices (blue) are the adjacency matrices for the inter edges. They are diagonal as each node is only linked to itself in the other layer. In the influence matrix used in the personalised multilayer PageRank algorithm, the common nodes represented with an X in the network are the source of influence (default).}
    \label{fig:ExMul}
\end{figure*}

\section{Multilayer Network Construction}\label{sec:MLN}

A graph, or single-layer network, $G=(V,E)$ consists of a set of nodes $V$ and a set of edges $E\subseteq V \times V$ that connect pairs of nodes \cite{newman2018networks}.  A network is represented by its rank-2 adjacency tensor $W_j^i$ which gives the intensity  --or weight-- of the link between nodes $i$ and $j$. The rank-2 adjacency tensor is more commonly called adjacency matrix. If the network has $N$ nodes then the dimension of $W_j^i$ is $N \times N$.

A bipartite network $B=(V_1,V_2,E)$ is a graph where the nodes belong to two disjoint sets, $V_1$ and $V_2$ and each edge connects a node in $V_1$ to a node in $V_2$ \cite{newman2018networks}. Let $N_1$ and $N_2$ be the number of nodes in each set. The adjacency matrix of a bipartite network is usually an $N_1 \times N_2$ matrix with entries representing edges between nodes in set $V_1$ (rows) and set $V_2$ (columns). However, in this paper we expand it to an $(N_1+ N_2)\times(N_1+N_2)$ matrix, so that it is similar to a regular graph. The entries in the upper $N_1\times N_1$ and lower $N_2\times N_2$ part are zero, to indicate that there are no edges between nodes in the same set. 

When there are more than one type of relationships between pairs of nodes in a network, we talk about multilayer networks. The nodes in each layer are connected based on one type of relationship, so the multiple layers are used to account for all different types of relationships. In this case there can be an edge between any pair of nodes in the same layer, or between any pair of nodes in any pair of layers.  These are called intra and inter edges, respectively.

In this paper we consider multilayer networks where in each layer there is a bipartite network, that is, a network with two types of nodes, and no edges between nodes of the same type.  The bipartite networks in each layer share a set of \emph{common nodes} (the borrowers) but the non-borrower type of nodes in each layer are different. We call these \emph{specific nodes}.

Now let $N$ be the total number of nodes in the network, that is, the sum of the number of common nodes, $N_c$, and the number of specific nodes in each layer. We use $V$ to denote the set of all nodes.
The subfigure on the left of Figure~\ref{fig:ExMul} shows an example of such a multilayer network with two layers. There are seven common nodes (black), two nodes which only belong to the bottom layer (brown) and three nodes which only belong to the top (red, green or yellow).

Let $\mathcal{N}$ be a multilayer network with $N$ nodes and $L$ layers. It can be represented by an $\mathbb{R}^{N\times N\times L\times L}$ dimensional rank-4 adjacency tensor $M_{j\beta}^{i\alpha}$ which indicates an edge between node $i$ in layer $\alpha$ and node $j$ in layer $\beta$ \cite{de2013mathematical}. By convention, nodes are represented with latin letters and layers with greek letters.

An equivalent representation is obtained by flattening $M_{j\beta}^{i\alpha}$ to obtain a rank-2 tensor which is an $N \cdot L \times N \cdot L$ matrix, known as supra adjacency tensor or matrix. Using this format, computations and notation are greatly simplified. 

If we assume, without loss of generality, that there are two layers, $\alpha$ and $\beta$, with intra layer adjacency matrices $A$ and $B$, $N_c$ common nodes, $N_{\alpha}$ and $N_{\beta}$ nodes related to specific layers $\alpha$ and $\beta$ respectively, then $M_{j\beta}^{i\alpha}$ can be written as
\[
M_{j\beta}^{i\alpha}=\left [\begin{array}{c|c}
    A_j^i & I \\ \hline
    I & B_j^i
\end{array} \right]
\]
where I is an $N\times N$ matrix with 1 on the diagonal for the first $N_c$ elements. Note that for computational convenience we include the specific nodes of all layers in the intra layer adjacency matrices, even though some edges are not allowed. The middle subfigure in Fig. \ref{fig:ExMul} shows the flattened adjacency tensor for the multilayer network on the left.

The supra transition matrix $T_{j\beta}^{i\alpha}$ is defined elementwise as $t_{lk}=\frac{m_{lk}}{\sum_{h=1}^{NL} m_{hk}}$ where $m_{lk}\in M_{j\beta}^{i\alpha}$, i.e. it is the column-normalised supra adjacency matrix, and it describes the transition probabilities of a random walker traversing the nodes within and between layers of the network \citep{garas2016interconnected}.

\section{Personalised Multilayer PageRank Centrality}\label{sec:PMP}
\subsection{Multilayer PageRank Centrality}\label{sec:multiplexpagerankc}
The PageRank centrality of a node is equivalent to the probability that a random walker who is traversing the network would end up at the node. The PageRank centrality in a single-layer network is the steady-state solution of the equation $p_j(t+1)=R_j^ip_i(t)$ where $R_i^j$ is the transition matrix of a random walk on the network, $r$ is the likelihood of moving along one of the edges of the current node to reach a neighboring node and $1-r$ is the likelihood of jumping --or teleporting-- to some other node in the network. The parameter $r$ is also called restart probability, and is usually set equal to 0.85 \cite{langville2004deeper}.
In a multilayer network, the jump can happen for any node in any layer.
The transition tensor for the random walk is therefore
\begin{equation}\label{eq:R}
    R^{i\alpha}_{j\beta}=rT^{i\alpha}_{j\beta}+\frac{1-r}{N \cdot L}u^{i\alpha}_{j\beta},
\end{equation}
where $T^{i\alpha}_{j\beta}$ is the rank-2 supra transition tensor, $r$ is the restart probability and $u^{i\alpha}_{j\beta}$ is an $N \cdot L \times N \cdot L$ matrix of ones.
This specification indicates that the random walk is equally likely to jump to any node in any layer of the multilayer network. 
This definition is valid for multilayer networks where all nodes have outgoing edges.

The transition tensor $R^{i\alpha}_{j\beta}$ represents the probability of jumping between pairs of nodes and switching between layers in the random walk. If $p_{i\alpha}(t)$ is the time dependent tensor that gives the probability of finding the walker in a given node and layer, then the random walk is described by
\[
p_{j\beta}(t+1)=R^{i\alpha}_{j\beta}p_{i\alpha}(t).
\]
To obtain the multilayer PageRank centrality, we find the steady state solution to this equation, that is, when $t \to \infty$. The solution is given by $\Pi_{i\alpha}$, which represents the probability of finding the walker at node $i$ in layer $\alpha$. The solution can be obtained by calculating the leading eigentensor, or the solution of the higher-order eigenvalue problem,
\[
T_{j\beta}^{i\alpha}\Pi_{i\alpha}=\lambda\Pi_{j\beta}
\]
as derived in \citet{domenico2013centrality}.

The steady state solution $\Omega_{i\alpha}$ gives the probability of finding the walker in node $i$ in layer $\alpha$ and in order to obtain a PageRank centrality for each node, the values in different layers are summed up, resulting in $\omega_i$ the multilayer PageRank centrality score for node $i$.

\subsection{Influence Matrix}
The multilayer PageRank centrality described above gives a representation of node importance assuming that a random walker jumps to any node in any layer with equal probability.
In many applications it is helpful to bring out nodes that are central from the perspective of a set of specific nodes. This is achieved by making the random walk biased, allowing the walker to jump to a set of specific nodes only. This is also called personalised PagerRank \cite{page1999pagerank} and has for example been applied to detect social security fraud with great success \cite{van2017gotcha}.

In order to personalise the PageRank for multilayer network, in Eq. \ref{eq:R} we modify the $u^{i\alpha}_{j\beta}$ matrix, which we call \emph{influence matrix}. Instead of setting all values to 1, to indicate an equal probability of the walker jumping to any node in any layer of the multilayer network, we only allow it to jump to certain nodes, which we call influence nodes, $V_I\in V$: the defaulters in our applications.

As before, $u^{i\alpha}_{j\beta}$ is an $N \cdot L\times N \cdot L$ matrix. Without loss of generality, we assume there are two layers, so that $u$ has dimension $2N\times 2N$, that can be split into four blocks of $N\times N$ sub-matrices. Two of those blocks are on the diagonal, $u_{11}$ and $u_{22}$, and correspond to the intra-layer edges; and the other two blocks are off the diagonal, in the upper and lower triangle, and correspond to the inter-layer edges. We call them $u_{12}$ and $u_{21}$. Thus
\[
u^{i\alpha}_{j\beta}=\left [\begin{array}{c|c}
    u_{11} & u_{12} \\ \hline
    u_{21} & u_{22}
\end{array} \right]
\]

To personalise the PageRank measure, we start by setting all values to 0 except for those where influence originates.
As the influence nodes appear several times in $u^{i\alpha}_{j\beta}$, both within layers and between layers, we assign the value 1 to the $V_I$ nodes on the diagonal of all four submatrices in $u^{i\alpha}_{j\beta}$. The subfigure on the right in Figure \ref{fig:ExMul} shows an example of the influence matrix for the multilayer network on the left, where influence originates with the nodes that are marked with a black X.

We are now ready to compute the PageRank centrality as in section \ref{sec:multiplexpagerankc} except the denominator in \ref{eq:R} reduces to the sum of the elements in $u^{i\alpha}_{j\beta}$.

\section{Experimental Results}\label{sec:Experiments}
\begin{figure*}
    \centering
    \includegraphics[scale=0.63]{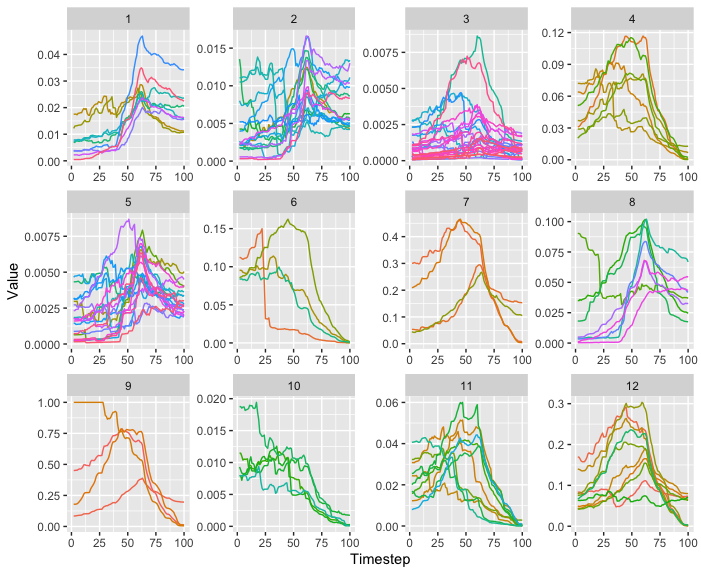}
    \caption{Product time series clusters. Each line represents the evolution of the influence (PageRank score) for each product node across time.}
    \label{fig:ProductClusters}
\end{figure*}

\begin{figure*}
    \centering
    \includegraphics[scale=0.63]{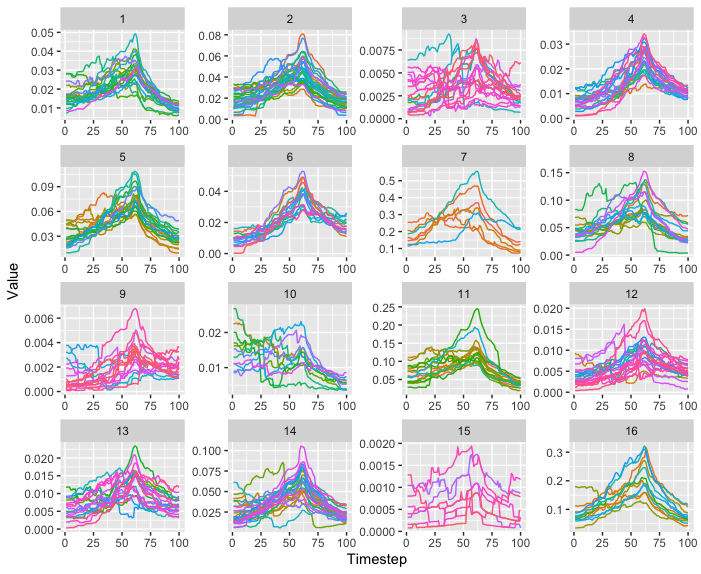}
    \caption{District time series clusters. Each line represents the evolution of the influence (PageRank score) for each district node across time.}
    \label{fig:DistrictCluster}
\end{figure*}
\subsection{Network Setup}
First, we start by describing our dataset. It corresponds to a subset of around 70,000 short-term agricultural loans, and it includes information on the loan itself (term, amount, guarantors), on the agribusiness requesting it (product, district they operate in) and finally past loan performance if available (number and amount of past loans, and any negative performance indicator). For more details regarding the data we refer the reader to \citet{Bravo_2013}. The data spans about 15 years and for each loan there is information about the year and month in which is was granted as well as whether the loan defaulted or not.

We build a sequence of multilayer networks, in which each network has two layers: one representing geographic location (district) and the other economic activity (product).  The common nodes represent the loans.  In the first layer, there are two types of nodes, loans and districts. Each loan is connected to the district in which the borrower operates in. In the example in Figure \ref{fig:ExMul}, there are two districts in the bottom layer denoted by brown nodes and seven borrowers connected to them (black nodes).

In the second layer, there are also two types of nodes, the loans and the products which the borrowers offer. Again, each loan is connected to its respective product. In the top layer of the network in Figure \ref{fig:ExMul} there are three product nodes, the red one could for example represent cherries, the green one avocado and the yellow one lemons.
The two layers are connected by creating an edge between the nodes representing the same borrower in each layer, indicated by dashed lines in the network in Figure~\ref{fig:ExMul}.

Each multilayer network therefore consists of two bipartite networks, and they are constructed using five years of loans. To  get subsequent networks we shift the time period by one month in each step. In total we have 100 such timesteps.

To each network in the sequence we apply the personalised multilayer PageRank algorithm derived in Section \ref{sec:PMP}, with influence originating at the nodes which represent loans that defaulted in the period. Thus we compute scores, --or default risk-- for all the nodes in the network. Therefore, in each timestep, we get a score for each loan, each product and each district. As we have 100 timesteps, this means we obtain a time series of length 100 for each product and district in the dataset. The scores represent credit risk exposure.

Overall, there are time series for 125 products and 276 districts. These time series describe the evolution of credit risk. 

\subsection{Evolution of Default Risk}
\begin{figure*}
    \centering
    \includegraphics[scale=0.4]{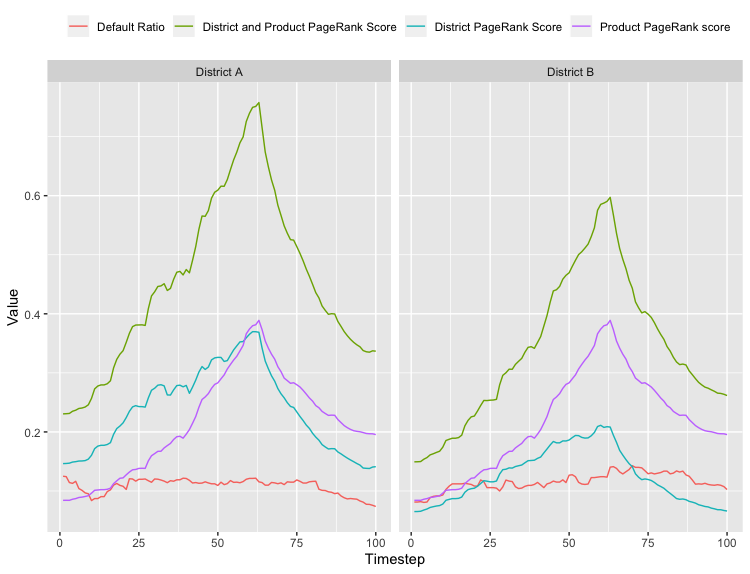}
    \caption{Evolution of risk for firewood in two districts (A and B). The figure shows the multilayer personalised PageRank scores for products, district, and the sum of product and district, together with the evolution of default rate of the product within the district.  }
    \label{fig:NetworkVsDefault}
\end{figure*}
\noindent \textbf{Clustering of Time Series}

\vspace{0.1cm}
Our first step to understand the evolution of risk comes from identifying common patterns in the time series induced by, on the one hand, product nodes and, on the other hand, district nodes. As the influence score represents the default influence that particular product has over the network at that time (normalised so at every timestep the total influence is constant), the evolution of the value of the influence for a given element at a particular time shows how default correlation changes across segments of farmers. 

We perform time series clustering to partition them into groups based on similarity because we want to identify products and districts that have similar risk evolution. Thereby we consider time series clustering by shape \cite{aghabozorgi2015time}, which attempts to match time series which are similar in both magnitude and trend. By clustering, we can observe how different products or districts behave similarly from a network point of view, as opposed to an overall risk trend which would be inferrable from a regression coefficient, for example.

As a distance metric, we use dynamic time warping, in order to find the optimal alignment between any pair of time series. Dynamic time warping is commonly used to find whole time series which are similar in shape \cite{aghabozorgi2015time}.
To cluster the time series we use the k-means clustering algorithm which is a partitioning method that makes $k$ clusters by minimising the total distance between all objects in a cluster and the cluster centre \citep{macqueen1967some}. 

Figures \ref{fig:ProductClusters} and \ref{fig:DistrictCluster} show the results of the clustering exercise for products and districts, respectively. Using the elbow method we found  that in the first case, the  optimal number of clusters was 12, while it was 16 in the second case. We can see how in most clusters the products and districts follow a similar trend. For example, cluster 10 in Figure~\ref{fig:ProductClusters} has four products (cattle, potatoes, fences, strawberry) whose influence scores trends down as time goes by. This means the propagation of risk across those products is less meaningful every month, so other factors (either individual or related to other layers of the network) should be the focus of risk mitigation actions. The reverse is true in cluster 1 for Figure~\ref{fig:ProductClusters}, where the risk of nine products (lemons, lettuce, corn, peach, tomatoes, blueberries) increases as time passes. 

We also point out the difference between clusters 4 and 12 in Figure \ref{fig:ProductClusters}. Although the time series in each cluster have a similar shape, the values themselves are much higher in cluster 12 than in cluster 4, indicating that overall, these products are more risky for lenders. It is noteworthy that most of the district time series in Figure \ref{fig:DistrictCluster} peak around the 60th timestep. These are the years 2000-2005 which corresponds to the end of the boom growth period experienced in the sector \cite{valdes2010}.

Now, a decision maker analysing the situation of customers with products in the latter, but operating in districts of the former knows the network-associated risk regarding the product is high while the district one is low. If the borrower has been flagged as high risk, targeted actions can be taken both towards them, but also less costly but blanket actions can be taken towards other borrowers which share the original borrowers' products as the network effects can hint at a risk factor that can be controlled and thus the overall risk can be diminished.

\vspace{0.05cm}
\noindent \textbf{Correlated Default versus Overall Trends}
\vspace{0.1cm}

The final experiment we conduct aims at verifying whether the influence score is reflecting network-related risk, or simply reflecting the default risk of a product, a district, or both jointly. For this, we studied the pairs [district, product] and calculated the independent influence scores, the sum of the scores (which represents the influence of the pair) and the default rate for the pair. The conclusion to be drawn is that the effects are network related. Default rate is slightly correlated with the network effects, but notably it shows a lag on its movements and it also presents lesser variability than network scores.

There are thousands of district and product pairs, so we show two representative examples in Figure~\ref{fig:NetworkVsDefault}. We see how the network risk increases as time passes, but in the second district the network risk quickly goes down. The default rates present a similar behaviour (default rate trending down at the end after a slight increase), but they do not show the level of detail that the network variable shows.

These results allow characterising the network effects across as many layers as the user has interest in studying, not only these two presented here. The use of bipartite networks, one for borrowers and one per layer for each other effect of interest, can be expanded as much as the analyst wishes. The impact of this is however a much larger matrix, so a balance must be reached between sufficient complexity of the network and complexity of calculations. In our experiments two networks with 70,000 borrowers was very manageable, suggesting much larger scaling capacity.

\section{Conclusions}
In this paper we developed a personalised PageRank centrality measure for multilayer networks which we then applied to a large set of agricultural loans to study correlation of default.  Although there is no obvious network structure in the dataset, connecting loans to their products and districts in two different layers resulted in a fully connected network, that allowed us to investigate default correlation. This method can be expanded to any connected dataset, greatly expanding the usefulness of graph models.

The clustering exercise over the product and district nodes show how similar products and districts can influence different operators (correlated default) and have a relevant industry application: decision makers can use these results to treat groups of borrowers that are at a similar risk early, as the evolution of the time series are better low level signals and allow for risk-oriented segmentations.

Furthermore, the results from this paper are encouraging as they also suggest the multilayered networks provide valuable information regarding the propagation of default risk across networks. The network scores follow trends which are more detailed than the swings in default rate would hint, thus revealing the potential of these scores to capture information related to propagation of default across specific individuals or specific nodes.

As future work, the predictive impact of these scores is our next goal. The use of influence scores as predictive variables has been a useful step in network environments and it has been shown to improve prediction, so it is the natural step forward in this research.

\section*{Acknowledgements}
The first author acknowledges the support of the Natural Sciences and Engineering Research Council of Canada (NSERC) [Discovery Grant RGPIN-2020-07114]. This research was undertaken, in part, thanks to funding from the Canada Research Chairs program.


\begin{thebibliography}{}

\bibitem [\protect \citeauthoryear {%
Aghabozorgi%
, Shirkhorshidi%
\BCBL {}\ \BBA {} Wah%
}{%
Aghabozorgi%
\ \protect \BOthers {.}}{%
{\protect \APACyear {2015}}%
}]{%
aghabozorgi2015time}
\APACinsertmetastar {%
aghabozorgi2015time}%
\begin{APACrefauthors}%
Aghabozorgi, S.%
, Shirkhorshidi, A\BPBI S.%
\BCBL {}\ \BBA {} Wah, T\BPBI Y.%
\end{APACrefauthors}%
\unskip\
\newblock
\APACrefYearMonthDay{2015}{}{}.
\newblock
{\BBOQ}\APACrefatitle {Time-series clustering--a decade review} {Time-series
  clustering--a decade review}.{\BBCQ}
\newblock
\APACjournalVolNumPages{Information Systems}{53}{}{16--38}.
\PrintBackRefs{\CurrentBib}

\bibitem [\protect \citeauthoryear {%
Anderson%
}{%
Anderson%
}{%
{\protect \APACyear {2019}}%
}]{%
anderson2019}
\APACinsertmetastar {%
anderson2019}%
\begin{APACrefauthors}%
Anderson, R.%
\end{APACrefauthors}%
\unskip\
\newblock
\APACrefYear{2019}.
\newblock
\APACrefbtitle {Credit {{Intelligence}} \& {{Modelling}}: {{Many Paths}}
  through the {{Forest}}} {Credit {{Intelligence}} \& {{Modelling}}: {{Many
  Paths}} through the {{Forest}}}.
\newblock
\APACaddressPublisher{{South Africa}}{{Ryan Risk Analytics, Inc}}.
\PrintBackRefs{\CurrentBib}

\bibitem [\protect \citeauthoryear {%
Arr{\'a}iz%
, Bruhn%
\BCBL {}\ \BBA {} Stucchi%
}{%
Arr{\'a}iz%
\ \protect \BOthers {.}}{%
{\protect \APACyear {2017}}%
}]{%
arraiz2017}
\APACinsertmetastar {%
arraiz2017}%
\begin{APACrefauthors}%
Arr{\'a}iz, I.%
, Bruhn, M.%
\BCBL {}\ \BBA {} Stucchi, R.%
\end{APACrefauthors}%
\unskip\
\newblock
\APACrefYearMonthDay{2017}{}{}.
\newblock
{\BBOQ}\APACrefatitle {Psychometrics as a {{Tool}} to {{Improve Credit
  Information}}} {Psychometrics as a {{Tool}} to {{Improve Credit
  Information}}}.{\BBCQ}
\newblock
\APACjournalVolNumPages{The World Bank Economic Review}{30}{S1}{S67--S76}.
\PrintBackRefs{\CurrentBib}

\bibitem [\protect \citeauthoryear {%
Barab{\'a}si%
\ \protect \BOthers {.}}{%
Barab{\'a}si%
\ \protect \BOthers {.}}{%
{\protect \APACyear {2016}}%
}]{%
barabasi2016network}
\APACinsertmetastar {%
barabasi2016network}%
\begin{APACrefauthors}%
Barab{\'a}si, A\BHBI L.%
\BCBT {}\ \BOthersPeriod {.}
\end{APACrefauthors}%
\unskip\
\newblock
\APACrefYear{2016}.
\newblock
\APACrefbtitle {Network science} {Network science}.
\newblock
\APACaddressPublisher{Cambridge, United Kingdom}{Cambridge university press}.
\PrintBackRefs{\CurrentBib}

\bibitem [\protect \citeauthoryear {%
{Basel Committee on Banking Supervision}%
}{%
{Basel Committee on Banking Supervision}%
}{%
{\protect \APACyear {2005}}%
}]{%
BCBSExplanatory2005}
\APACinsertmetastar {%
BCBSExplanatory2005}%
\begin{APACrefauthors}%
{Basel Committee on Banking Supervision}.%
\end{APACrefauthors}%
\unskip\
\newblock
\APACrefYearMonthDay{2005}{{\APACmonth{07}}}{}.
\newblock
\APACrefbtitle {An {{Explanatory Note}} on the {{Basel II IRB Risk Weight
  Functions}}} {An {{Explanatory Note}} on the {{Basel II IRB Risk Weight
  Functions}}}\ \APACbVolEdTR{}{\BTR{}}.
\newblock
\APACaddressInstitution{}{{Bank for International Settlements}}.
\PrintBackRefs{\CurrentBib}

\bibitem [\protect \citeauthoryear {%
Botelho%
\ \BBA {} Antunes%
}{%
Botelho%
\ \BBA {} Antunes%
}{%
{\protect \APACyear {2011}}%
}]{%
botelho2011combining}
\APACinsertmetastar {%
botelho2011combining}%
\begin{APACrefauthors}%
Botelho, J.%
\BCBT {}\ \BBA {} Antunes, C.%
\end{APACrefauthors}%
\unskip\
\newblock
\APACrefYearMonthDay{2011}{}{}.
\newblock
{\BBOQ}\APACrefatitle {Combining Social Network Analysis with Semi-supervised
  Clustering: a case study on fraud detection} {Combining social network
  analysis with semi-supervised clustering: a case study on fraud
  detection}.{\BBCQ}
\newblock
\BIn{} \APACrefbtitle {Proceeding of Mining Data Semantics (MDS’2011) in
  Conjunction with SIGKDD} {Proceeding of mining data semantics (mds’2011) in
  conjunction with sigkdd}\ (\BPGS\ 1--7).
\newblock
\APACaddressPublisher{San Diego, CA, USA}{Citeseer}.
\PrintBackRefs{\CurrentBib}

\bibitem [\protect \citeauthoryear {%
Bravo%
, Maldonado%
\BCBL {}\ \BBA {} Weber%
}{%
Bravo%
\ \protect \BOthers {.}}{%
{\protect \APACyear {2013}}%
}]{%
Bravo_2013}
\APACinsertmetastar {%
Bravo_2013}%
\begin{APACrefauthors}%
Bravo, C.%
, Maldonado, S.%
\BCBL {}\ \BBA {} Weber, R.%
\end{APACrefauthors}%
\unskip\
\newblock
\APACrefYearMonthDay{2013}{}{}.
\newblock
{\BBOQ}\APACrefatitle {Granting and Managing Loans for Micro-Entrepreneurs:
  {{New}} Developments and Practical Experiences} {Granting and managing loans
  for micro-entrepreneurs: {{New}} developments and practical
  experiences}.{\BBCQ}
\newblock
\APACjournalVolNumPages{European Journal of Operational Research}{227}{2}{358 -
  366}.
\PrintBackRefs{\CurrentBib}

\bibitem [\protect \citeauthoryear {%
Calabrese%
, Andreeva%
\BCBL {}\ \BBA {} Ansell%
}{%
Calabrese%
\ \protect \BOthers {.}}{%
{\protect \APACyear {2019}}%
}]{%
calabrese2019}
\APACinsertmetastar {%
calabrese2019}%
\begin{APACrefauthors}%
Calabrese, R.%
, Andreeva, G.%
\BCBL {}\ \BBA {} Ansell, J.%
\end{APACrefauthors}%
\unskip\
\newblock
\APACrefYearMonthDay{2019}{}{}.
\newblock
{\BBOQ}\APACrefatitle {``{{Birds}} of a {{Feather}}'' {{Fail Together}}:
  {{Exploring}} the {{Nature}} of {{Dependency}} in {{SME Defaults}}}
  {``{{Birds}} of a {{Feather}}'' {{Fail Together}}: {{Exploring}} the
  {{Nature}} of {{Dependency}} in {{SME Defaults}}}.{\BBCQ}
\newblock
\APACjournalVolNumPages{Risk Analysis}{39}{1}{71-84}.
\PrintBackRefs{\CurrentBib}

\bibitem [\protect \citeauthoryear {%
Cheney%
}{%
Cheney%
}{%
{\protect \APACyear {2008}}%
}]{%
cheney2008}
\APACinsertmetastar {%
cheney2008}%
\begin{APACrefauthors}%
Cheney, J\BPBI S.%
\end{APACrefauthors}%
\unskip\
\newblock
\APACrefYearMonthDay{2008}{{\APACmonth{02}}}{}.
\newblock
\APACrefbtitle {Alternative {{Data}} and Its {{Use}} in {{Credit Scoring
  Thin}}- and {{No}}-{{File Consumers}}} {Alternative {{Data}} and its {{Use}}
  in {{Credit Scoring Thin}}- and {{No}}-{{File Consumers}}}\ \APACbVolEdTR
  {}{Payment {{Cards Center Discussion Paper}}\ \BNUM\ 08-01}.
\newblock
\APACaddressInstitution{{Philadelphia, USA}}{{Federal Reserve Bank of
  Philadelphia}}.
\PrintBackRefs{\CurrentBib}

\bibitem [\protect \citeauthoryear {%
De~Domenico%
, Porter%
\BCBL {}\ \BBA {} Arenas%
}{%
De~Domenico%
\ \protect \BOthers {.}}{%
{\protect \APACyear {2015}}%
}]{%
de2015muxviz}
\APACinsertmetastar {%
de2015muxviz}%
\begin{APACrefauthors}%
De~Domenico, M.%
, Porter, M\BPBI A.%
\BCBL {}\ \BBA {} Arenas, A.%
\end{APACrefauthors}%
\unskip\
\newblock
\APACrefYearMonthDay{2015}{}{}.
\newblock
{\BBOQ}\APACrefatitle {MuxViz: a tool for multilayer analysis and visualization
  of networks} {Muxviz: a tool for multilayer analysis and visualization of
  networks}.{\BBCQ}
\newblock
\APACjournalVolNumPages{Journal of Complex Networks}{3}{2}{159--176}.
\PrintBackRefs{\CurrentBib}

\bibitem [\protect \citeauthoryear {%
De~Domenico%
\ \protect \BOthers {.}}{%
De~Domenico%
\ \protect \BOthers {.}}{%
{\protect \APACyear {2013}}%
}]{%
de2013mathematical}
\APACinsertmetastar {%
de2013mathematical}%
\begin{APACrefauthors}%
De~Domenico, M.%
, Sol{\'e}-Ribalta, A.%
, Cozzo, E.%
, Kivel{\"a}, M.%
, Moreno, Y.%
, Porter, M\BPBI A.%
\BDBL {}Arenas, A.%
\end{APACrefauthors}%
\unskip\
\newblock
\APACrefYearMonthDay{2013}{}{}.
\newblock
{\BBOQ}\APACrefatitle {Mathematical formulation of multilayer networks}
  {Mathematical formulation of multilayer networks}.{\BBCQ}
\newblock
\APACjournalVolNumPages{Physical Review X}{3}{4}{041022}.
\PrintBackRefs{\CurrentBib}

\bibitem [\protect \citeauthoryear {%
Domenico%
, Sol-Ribalta%
, Omodei%
, Gmez%
\BCBL {}\ \BBA {} Arenas%
}{%
Domenico%
\ \protect \BOthers {.}}{%
{\protect \APACyear {2015}}%
}]{%
domenico2013centrality}
\APACinsertmetastar {%
domenico2013centrality}%
\begin{APACrefauthors}%
Domenico, M.%
, Sol-Ribalta, A.%
, Omodei, E.%
, Gmez, S.%
\BCBL {}\ \BBA {} Arenas, A.%
\end{APACrefauthors}%
\unskip\
\newblock
\APACrefYearMonthDay{2015}{}{}.
\newblock
{\BBOQ}\APACrefatitle {Ranking in interconnected multilayer networks reveals
  versatile nodes} {Ranking in interconnected multilayer networks reveals
  versatile nodes}.{\BBCQ}
\newblock
\APACjournalVolNumPages{Nature Communications}{6}{}{6868}.
\PrintBackRefs{\CurrentBib}

\bibitem [\protect \citeauthoryear {%
Duan%
\ \BBA {} Miao%
}{%
Duan%
\ \BBA {} Miao%
}{%
{\protect \APACyear {2016}}%
}]{%
duan2016}
\APACinsertmetastar {%
duan2016}%
\begin{APACrefauthors}%
Duan, J\BHBI C.%
\BCBT {}\ \BBA {} Miao, W.%
\end{APACrefauthors}%
\unskip\
\newblock
\APACrefYearMonthDay{2016}{{\APACmonth{10}}}{}.
\newblock
{\BBOQ}\APACrefatitle {Default {{Correlations}} and {{Large}}-{{Portfolio
  Credit Analysis}}} {Default {{Correlations}} and {{Large}}-{{Portfolio Credit
  Analysis}}}.{\BBCQ}
\newblock
\APACjournalVolNumPages{Journal of Business \& Economic
  Statistics}{34}{4}{536-546}.
\PrintBackRefs{\CurrentBib}

\bibitem [\protect \citeauthoryear {%
Fenech%
, Vosgha%
\BCBL {}\ \BBA {} Shafik%
}{%
Fenech%
\ \protect \BOthers {.}}{%
{\protect \APACyear {2015}}%
}]{%
fenech2015}
\APACinsertmetastar {%
fenech2015}%
\begin{APACrefauthors}%
Fenech, J\BPBI P.%
, Vosgha, H.%
\BCBL {}\ \BBA {} Shafik, S.%
\end{APACrefauthors}%
\unskip\
\newblock
\APACrefYearMonthDay{2015}{}{}.
\newblock
{\BBOQ}\APACrefatitle {Loan Default Correlation Using an {{Archimedean}} Copula
  Approach: {{A}} Case for Recalibration} {Loan default correlation using an
  {{Archimedean}} copula approach: {{A}} case for recalibration}.{\BBCQ}
\newblock
\APACjournalVolNumPages{Economic Modelling}{47}{}{340-354}.
\PrintBackRefs{\CurrentBib}

\bibitem [\protect \citeauthoryear {%
Garas%
}{%
Garas%
}{%
{\protect \APACyear {2016}}%
}]{%
garas2016interconnected}
\APACinsertmetastar {%
garas2016interconnected}%
\begin{APACrefauthors}%
Garas, A.%
\end{APACrefauthors}%
\unskip\
\newblock
\APACrefYear{2016}.
\newblock
\APACrefbtitle {Interconnected networks} {Interconnected networks}.
\newblock
\APACaddressPublisher{NY, USA}{Springer}.
\PrintBackRefs{\CurrentBib}

\bibitem [\protect \citeauthoryear {%
Gomez%
\ \protect \BOthers {.}}{%
Gomez%
\ \protect \BOthers {.}}{%
{\protect \APACyear {2013}}%
}]{%
gomez2013diffusion}
\APACinsertmetastar {%
gomez2013diffusion}%
\begin{APACrefauthors}%
Gomez, S.%
, Diaz-Guilera, A.%
, Gomez-Gardenes, J.%
, Perez-Vicente, C\BPBI J.%
, Moreno, Y.%
\BCBL {}\ \BBA {} Arenas, A.%
\end{APACrefauthors}%
\unskip\
\newblock
\APACrefYearMonthDay{2013}{}{}.
\newblock
{\BBOQ}\APACrefatitle {Diffusion dynamics on multiplex networks} {Diffusion
  dynamics on multiplex networks}.{\BBCQ}
\newblock
\APACjournalVolNumPages{Physical review letters}{110}{2}{028701}.
\PrintBackRefs{\CurrentBib}

\bibitem [\protect \citeauthoryear {%
Halu%
, Mondrag{\'o}n%
, Panzarasa%
\BCBL {}\ \BBA {} Bianconi%
}{%
Halu%
\ \protect \BOthers {.}}{%
{\protect \APACyear {2013}}%
}]{%
halu2013multiplex}
\APACinsertmetastar {%
halu2013multiplex}%
\begin{APACrefauthors}%
Halu, A.%
, Mondrag{\'o}n, R\BPBI J.%
, Panzarasa, P.%
\BCBL {}\ \BBA {} Bianconi, G.%
\end{APACrefauthors}%
\unskip\
\newblock
\APACrefYearMonthDay{2013}{}{}.
\newblock
{\BBOQ}\APACrefatitle {Multiplex pagerank} {Multiplex pagerank}.{\BBCQ}
\newblock
\APACjournalVolNumPages{PloS one}{8}{10}{e78293}.
\PrintBackRefs{\CurrentBib}

\bibitem [\protect \citeauthoryear {%
Iacovacci%
\ \BBA {} Bianconi%
}{%
Iacovacci%
\ \BBA {} Bianconi%
}{%
{\protect \APACyear {2016}}%
}]{%
iacovacci2016extracting}
\APACinsertmetastar {%
iacovacci2016extracting}%
\begin{APACrefauthors}%
Iacovacci, J.%
\BCBT {}\ \BBA {} Bianconi, G.%
\end{APACrefauthors}%
\unskip\
\newblock
\APACrefYearMonthDay{2016}{}{}.
\newblock
{\BBOQ}\APACrefatitle {Extracting information from multiplex networks}
  {Extracting information from multiplex networks}.{\BBCQ}
\newblock
\APACjournalVolNumPages{Chaos: An Interdisciplinary Journal of Nonlinear
  Science}{26}{6}{065306}.
\PrintBackRefs{\CurrentBib}

\bibitem [\protect \citeauthoryear {%
Iacovacci%
, Rahmede%
, Arenas%
\BCBL {}\ \BBA {} Bianconi%
}{%
Iacovacci%
\ \protect \BOthers {.}}{%
{\protect \APACyear {2016}}%
}]{%
iacovacci2016functional}
\APACinsertmetastar {%
iacovacci2016functional}%
\begin{APACrefauthors}%
Iacovacci, J.%
, Rahmede, C.%
, Arenas, A.%
\BCBL {}\ \BBA {} Bianconi, G.%
\end{APACrefauthors}%
\unskip\
\newblock
\APACrefYearMonthDay{2016}{}{}.
\newblock
{\BBOQ}\APACrefatitle {Functional multiplex pagerank} {Functional multiplex
  pagerank}.{\BBCQ}
\newblock
\APACjournalVolNumPages{EPL (Europhysics Letters)}{116}{2}{28004}.
\PrintBackRefs{\CurrentBib}

\bibitem [\protect \citeauthoryear {%
Kivel{\"a}%
\ \protect \BOthers {.}}{%
Kivel{\"a}%
\ \protect \BOthers {.}}{%
{\protect \APACyear {2014}}%
}]{%
kivela2014multilayer}
\APACinsertmetastar {%
kivela2014multilayer}%
\begin{APACrefauthors}%
Kivel{\"a}, M.%
, Arenas, A.%
, Barthelemy, M.%
, Gleeson, J\BPBI P.%
, Moreno, Y.%
\BCBL {}\ \BBA {} Porter, M\BPBI A.%
\end{APACrefauthors}%
\unskip\
\newblock
\APACrefYearMonthDay{2014}{}{}.
\newblock
{\BBOQ}\APACrefatitle {Multilayer networks} {Multilayer networks}.{\BBCQ}
\newblock
\APACjournalVolNumPages{Journal of complex networks}{2}{3}{203--271}.
\PrintBackRefs{\CurrentBib}

\bibitem [\protect \citeauthoryear {%
Kwak%
, Lee%
, Park%
\BCBL {}\ \BBA {} Moon%
}{%
Kwak%
\ \protect \BOthers {.}}{%
{\protect \APACyear {2010}}%
}]{%
kwak2010twitter}
\APACinsertmetastar {%
kwak2010twitter}%
\begin{APACrefauthors}%
Kwak, H.%
, Lee, C.%
, Park, H.%
\BCBL {}\ \BBA {} Moon, S.%
\end{APACrefauthors}%
\unskip\
\newblock
\APACrefYearMonthDay{2010}{}{}.
\newblock
{\BBOQ}\APACrefatitle {What is Twitter, a social network or a news media?}
  {What is twitter, a social network or a news media?}{\BBCQ}
\newblock
\BIn{} \APACrefbtitle {Proceedings of the 19th international conference on
  World wide web} {Proceedings of the 19th international conference on world
  wide web}\ (\BPGS\ 591--600).
\newblock
\APACaddressPublisher{Raleigh, NC, USA}{ACM}.
\PrintBackRefs{\CurrentBib}

\bibitem [\protect \citeauthoryear {%
Langville%
\ \BBA {} Meyer%
}{%
Langville%
\ \BBA {} Meyer%
}{%
{\protect \APACyear {2004}}%
}]{%
langville2004deeper}
\APACinsertmetastar {%
langville2004deeper}%
\begin{APACrefauthors}%
Langville, A\BPBI N.%
\BCBT {}\ \BBA {} Meyer, C\BPBI D.%
\end{APACrefauthors}%
\unskip\
\newblock
\APACrefYearMonthDay{2004}{}{}.
\newblock
{\BBOQ}\APACrefatitle {Deeper inside pagerank} {Deeper inside pagerank}.{\BBCQ}
\newblock
\APACjournalVolNumPages{Internet Mathematics}{1}{3}{335--380}.
\PrintBackRefs{\CurrentBib}

\bibitem [\protect \citeauthoryear {%
Lessmann%
, Baesens%
, Seow%
\BCBL {}\ \BBA {} Thomas%
}{%
Lessmann%
\ \protect \BOthers {.}}{%
{\protect \APACyear {2015}}%
}]{%
lessmann2015}
\APACinsertmetastar {%
lessmann2015}%
\begin{APACrefauthors}%
Lessmann, S.%
, Baesens, B.%
, Seow, H\BHBI V.%
\BCBL {}\ \BBA {} Thomas, L\BPBI C.%
\end{APACrefauthors}%
\unskip\
\newblock
\APACrefYearMonthDay{2015}{{\APACmonth{11}}}{}.
\newblock
{\BBOQ}\APACrefatitle {Benchmarking State-of-the-Art Classification Algorithms
  for Credit Scoring: {{An}} Update of Research} {Benchmarking state-of-the-art
  classification algorithms for credit scoring: {{An}} update of
  research}.{\BBCQ}
\newblock
\APACjournalVolNumPages{European Journal of Operational
  Research}{247}{1}{124-136}.
\PrintBackRefs{\CurrentBib}

\bibitem [\protect \citeauthoryear {%
Lohmann%
\ \protect \BOthers {.}}{%
Lohmann%
\ \protect \BOthers {.}}{%
{\protect \APACyear {2010}}%
}]{%
lohmann2010eigenvector}
\APACinsertmetastar {%
lohmann2010eigenvector}%
\begin{APACrefauthors}%
Lohmann, G.%
, Margulies, D\BPBI S.%
, Horstmann, A.%
, Pleger, B.%
, Lepsien, J.%
, Goldhahn, D.%
\BDBL {}Turner, R.%
\end{APACrefauthors}%
\unskip\
\newblock
\APACrefYearMonthDay{2010}{}{}.
\newblock
{\BBOQ}\APACrefatitle {Eigenvector centrality mapping for analyzing
  connectivity patterns in fMRI data of the human brain} {Eigenvector
  centrality mapping for analyzing connectivity patterns in fmri data of the
  human brain}.{\BBCQ}
\newblock
\APACjournalVolNumPages{PloS one}{5}{4}{e10232}.
\PrintBackRefs{\CurrentBib}

\bibitem [\protect \citeauthoryear {%
MacQueen%
\ \protect \BOthers {.}}{%
MacQueen%
\ \protect \BOthers {.}}{%
{\protect \APACyear {1967}}%
}]{%
macqueen1967some}
\APACinsertmetastar {%
macqueen1967some}%
\begin{APACrefauthors}%
MacQueen, J.%
\BCBT {}\ \BOthersPeriod {.}
\end{APACrefauthors}%
\unskip\
\newblock
\APACrefYearMonthDay{1967}{}{}.
\newblock
{\BBOQ}\APACrefatitle {Some methods for classification and analysis of
  multivariate observations} {Some methods for classification and analysis of
  multivariate observations}.{\BBCQ}
\newblock
\BIn{} \APACrefbtitle {Proceedings of the fifth Berkeley symposium on
  mathematical statistics and probability} {Proceedings of the fifth berkeley
  symposium on mathematical statistics and probability}\ (\BPGS\ 281--297).
\newblock
\APACaddressPublisher{Oakland, CA, USA}{University of California Press}.
\PrintBackRefs{\CurrentBib}

\bibitem [\protect \citeauthoryear {%
Mcauley%
\ \BBA {} Leskovec%
}{%
Mcauley%
\ \BBA {} Leskovec%
}{%
{\protect \APACyear {2014}}%
}]{%
mcauley2014}
\APACinsertmetastar {%
mcauley2014}%
\begin{APACrefauthors}%
Mcauley, J.%
\BCBT {}\ \BBA {} Leskovec, J.%
\end{APACrefauthors}%
\unskip\
\newblock
\APACrefYearMonthDay{2014}{{\APACmonth{02}}}{}.
\newblock
{\BBOQ}\APACrefatitle {Discovering {{Social Circles}} in {{Ego Networks}}}
  {Discovering {{Social Circles}} in {{Ego Networks}}}.{\BBCQ}
\newblock
\APACjournalVolNumPages{ACM Trans. Knowl. Discov. Data}{8}{1}{4:1--4:28}.
\PrintBackRefs{\CurrentBib}

\bibitem [\protect \citeauthoryear {%
Min%
\ \protect \BOthers {.}}{%
Min%
\ \protect \BOthers {.}}{%
{\protect \APACyear {2018}}%
}]{%
min2018behavior}
\APACinsertmetastar {%
min2018behavior}%
\begin{APACrefauthors}%
Min, W.%
, Tang, Z.%
, Zhu, M.%
, Dai, Y.%
, Wei, Y.%
\BCBL {}\ \BBA {} Zhang, R.%
\end{APACrefauthors}%
\unskip\
\newblock
\APACrefYearMonthDay{2018}{}{}.
\newblock
{\BBOQ}\APACrefatitle {Behavior Language Processing with Graph based Feature
  Generation for Fraud Detection in Online Lending} {Behavior language
  processing with graph based feature generation for fraud detection in online
  lending}.{\BBCQ}
\newblock
\BIn{} \APACrefbtitle {Proceedings of Workshop on Misinformation and
  Misbehavior Mining on the Web} {Proceedings of workshop on misinformation and
  misbehavior mining on the web}\ (\BPGS\ 1--8).
\newblock
\APACaddressPublisher{Marina Del Rey, CA, USA}{ACM}.
\PrintBackRefs{\CurrentBib}

\bibitem [\protect \citeauthoryear {%
Newman%
}{%
Newman%
}{%
{\protect \APACyear {2018}}%
}]{%
newman2018networks}
\APACinsertmetastar {%
newman2018networks}%
\begin{APACrefauthors}%
Newman, M.%
\end{APACrefauthors}%
\unskip\
\newblock
\APACrefYear{2018}.
\newblock
\APACrefbtitle {Networks} {Networks}.
\newblock
\APACaddressPublisher{Oxford, UK}{Oxford University Press}.
\PrintBackRefs{\CurrentBib}

\bibitem [\protect \citeauthoryear {%
{\'O}skarsd{\'o}ttir%
, Bravo%
, Sarraute%
, Vanthienen%
\BCBL {}\ \BBA {} Baesens%
}{%
{\'O}skarsd{\'o}ttir%
\ \protect \BOthers {.}}{%
{\protect \APACyear {2019}}%
}]{%
oskarsdottir2019}
\APACinsertmetastar {%
oskarsdottir2019}%
\begin{APACrefauthors}%
{\'O}skarsd{\'o}ttir, M.%
, Bravo, C.%
, Sarraute, C.%
, Vanthienen, J.%
\BCBL {}\ \BBA {} Baesens, B.%
\end{APACrefauthors}%
\unskip\
\newblock
\APACrefYearMonthDay{2019}{}{}.
\newblock
{\BBOQ}\APACrefatitle {The value of big data for credit scoring: Enhancing
  financial inclusion using mobile phone data and social network analytics}
  {The value of big data for credit scoring: Enhancing financial inclusion
  using mobile phone data and social network analytics}.{\BBCQ}
\newblock
\APACjournalVolNumPages{Applied Soft Computing}{74}{}{26--39}.
\PrintBackRefs{\CurrentBib}

\bibitem [\protect \citeauthoryear {%
Page%
, Brin%
, Motwani%
\BCBL {}\ \BBA {} Winograd%
}{%
Page%
\ \protect \BOthers {.}}{%
{\protect \APACyear {1999}}%
}]{%
page1999pagerank}
\APACinsertmetastar {%
page1999pagerank}%
\begin{APACrefauthors}%
Page, L.%
, Brin, S.%
, Motwani, R.%
\BCBL {}\ \BBA {} Winograd, T.%
\end{APACrefauthors}%
\unskip\
\newblock
\APACrefYearMonthDay{1999}{}{}.
\newblock
\APACrefbtitle {The pagerank citation ranking: Bringing order to the web.} {The
  pagerank citation ranking: Bringing order to the web.}\
  \APACbVolEdTR{}{\BTR{}}.
\newblock
\APACaddressInstitution{}{Stanford InfoLab}.
\PrintBackRefs{\CurrentBib}

\bibitem [\protect \citeauthoryear {%
Pedroche%
, Romance%
\BCBL {}\ \BBA {} Criado%
}{%
Pedroche%
\ \protect \BOthers {.}}{%
{\protect \APACyear {2016}}%
}]{%
pedroche2016biplex}
\APACinsertmetastar {%
pedroche2016biplex}%
\begin{APACrefauthors}%
Pedroche, F.%
, Romance, M.%
\BCBL {}\ \BBA {} Criado, R.%
\end{APACrefauthors}%
\unskip\
\newblock
\APACrefYearMonthDay{2016}{}{}.
\newblock
{\BBOQ}\APACrefatitle {A biplex approach to PageRank centrality: From classic
  to multiplex networks} {A biplex approach to pagerank centrality: From
  classic to multiplex networks}.{\BBCQ}
\newblock
\APACjournalVolNumPages{Chaos: An Interdisciplinary Journal of Nonlinear
  Science}{26}{6}{065301}.
\PrintBackRefs{\CurrentBib}

\bibitem [\protect \citeauthoryear {%
Poledna%
, Molina-Borboa%
, Mart{\'\i}nez-Jaramillo%
, Van Der~Leij%
\BCBL {}\ \BBA {} Thurner%
}{%
Poledna%
\ \protect \BOthers {.}}{%
{\protect \APACyear {2015}}%
}]{%
poledna2015multi}
\APACinsertmetastar {%
poledna2015multi}%
\begin{APACrefauthors}%
Poledna, S.%
, Molina-Borboa, J\BPBI L.%
, Mart{\'\i}nez-Jaramillo, S.%
, Van Der~Leij, M.%
\BCBL {}\ \BBA {} Thurner, S.%
\end{APACrefauthors}%
\unskip\
\newblock
\APACrefYearMonthDay{2015}{}{}.
\newblock
{\BBOQ}\APACrefatitle {The multi-layer network nature of systemic risk and its
  implications for the costs of financial crises} {The multi-layer network
  nature of systemic risk and its implications for the costs of financial
  crises}.{\BBCQ}
\newblock
\APACjournalVolNumPages{Journal of Financial Stability}{20}{}{70--81}.
\PrintBackRefs{\CurrentBib}

\bibitem [\protect \citeauthoryear {%
Renton%
}{%
Renton%
}{%
{\protect \APACyear {2019}}%
}]{%
renton2019}
\APACinsertmetastar {%
renton2019}%
\begin{APACrefauthors}%
Renton, P.%
\end{APACrefauthors}%
\unskip\
\newblock
\APACrefYearMonthDay{2019}{{\APACmonth{11}}}{}.
\newblock
\APACrefbtitle {Experian {{Introduces New Credit Scoring Suite}} for {{Thin
  File Consumers}}.} {Experian {{Introduces New Credit Scoring Suite}} for
  {{Thin File Consumers}}.}
\newblock
\APAChowpublished
  {https://www.lendacademy.com/experian-introduces-new-credit-scoring-suite-for-thin-file-consumers/}.
\PrintBackRefs{\CurrentBib}

\bibitem [\protect \citeauthoryear {%
{San Pedro}%
, Proserpio%
\BCBL {}\ \BBA {} Oliver%
}{%
{San Pedro}%
\ \protect \BOthers {.}}{%
{\protect \APACyear {2015}}%
}]{%
sanpedro2015}
\APACinsertmetastar {%
sanpedro2015}%
\begin{APACrefauthors}%
{San Pedro}, J.%
, Proserpio, D.%
\BCBL {}\ \BBA {} Oliver, N.%
\end{APACrefauthors}%
\unskip\
\newblock
\APACrefYearMonthDay{2015}{}{}.
\newblock
{\BBOQ}\APACrefatitle {{{MobiScore}}: {{Towards Universal Credit Scoring}} from
  {{Mobile Phone Data}}} {{{MobiScore}}: {{Towards Universal Credit Scoring}}
  from {{Mobile Phone Data}}}.{\BBCQ}
\newblock
\BIn{} F.~Ricci, K.~Bontcheva, O.~Conlan\BCBL {}\ \BBA {} S.~Lawless\ (\BEDS),
  \APACrefbtitle {User {{Modeling}}, {{Adaptation}} and {{Personalization}}}
  {User {{Modeling}}, {{Adaptation}} and {{Personalization}}}\ (\BPG~195-207).
\newblock
\APACaddressPublisher{{Cham}}{{Springer International Publishing}}.
\PrintBackRefs{\CurrentBib}

\bibitem [\protect \citeauthoryear {%
Siddiqi%
}{%
Siddiqi%
}{%
{\protect \APACyear {2017}}%
}]{%
Siddiqi2017}
\APACinsertmetastar {%
Siddiqi2017}%
\begin{APACrefauthors}%
Siddiqi, N.%
\end{APACrefauthors}%
\unskip\
\newblock
\APACrefYear{2017}.
\newblock
\APACrefbtitle {Intelligent Credit Scoring: Building and Implementing Better
  Credit Risk Scorecard} {Intelligent credit scoring: Building and implementing
  better credit risk scorecard}.
\newblock
\APACaddressPublisher{NY, USA}{{John Wiley {{\&}} Sons, Inc.}}
\PrintBackRefs{\CurrentBib}

\bibitem [\protect \citeauthoryear {%
L.~Thomas%
, Crook%
\BCBL {}\ \BBA {} Edelman%
}{%
L.~Thomas%
\ \protect \BOthers {.}}{%
{\protect \APACyear {2017}}%
}]{%
thomas2017}
\APACinsertmetastar {%
thomas2017}%
\begin{APACrefauthors}%
Thomas, L.%
, Crook, J.%
\BCBL {}\ \BBA {} Edelman, D.%
\end{APACrefauthors}%
\unskip\
\newblock
\APACrefYear{2017}.
\newblock
\APACrefbtitle {Credit {{Scoring}} and {{Its Applications}}, {{Second
  Edition}}} {Credit {{Scoring}} and {{Its Applications}}, {{Second Edition}}}\
  (\PrintOrdinal{Second Edition}\ \BEd).
\newblock
\APACaddressPublisher{{USA}}{{SIAM}}.
\PrintBackRefs{\CurrentBib}

\bibitem [\protect \citeauthoryear {%
L\BPBI C.~Thomas%
, Oliver%
\BCBL {}\ \BBA {} Hand%
}{%
L\BPBI C.~Thomas%
\ \protect \BOthers {.}}{%
{\protect \APACyear {2005}}%
}]{%
Thomas_2005}
\APACinsertmetastar {%
Thomas_2005}%
\begin{APACrefauthors}%
Thomas, L\BPBI C.%
, Oliver, R\BPBI W.%
\BCBL {}\ \BBA {} Hand, D\BPBI J.%
\end{APACrefauthors}%
\unskip\
\newblock
\APACrefYearMonthDay{2005}{}{}.
\newblock
{\BBOQ}\APACrefatitle {A Survey of the Issues in Consumer Credit Modelling
  Research} {A survey of the issues in consumer credit modelling
  research}.{\BBCQ}
\newblock
\APACjournalVolNumPages{The Journal of the Operational Research
  Society}{56}{9}{1006-1015}.
\PrintBackRefs{\CurrentBib}

\bibitem [\protect \citeauthoryear {%
Thurner%
\ \BBA {} Poledna%
}{%
Thurner%
\ \BBA {} Poledna%
}{%
{\protect \APACyear {2013}}%
}]{%
thurner2013debtrank}
\APACinsertmetastar {%
thurner2013debtrank}%
\begin{APACrefauthors}%
Thurner, S.%
\BCBT {}\ \BBA {} Poledna, S.%
\end{APACrefauthors}%
\unskip\
\newblock
\APACrefYearMonthDay{2013}{}{}.
\newblock
{\BBOQ}\APACrefatitle {DebtRank-transparency: Controlling systemic risk in
  financial networks} {Debtrank-transparency: Controlling systemic risk in
  financial networks}.{\BBCQ}
\newblock
\APACjournalVolNumPages{Scientific reports}{3}{}{1888}.
\PrintBackRefs{\CurrentBib}

\bibitem [\protect \citeauthoryear {%
Vald{\'e}s%
, Foster%
, P{\'e}rez%
\BCBL {}\ \BBA {} Rivera%
}{%
Vald{\'e}s%
\ \protect \BOthers {.}}{%
{\protect \APACyear {2010}}%
}]{%
valdes2010}
\APACinsertmetastar {%
valdes2010}%
\begin{APACrefauthors}%
Vald{\'e}s, A.%
, Foster, W.%
, P{\'e}rez, R.%
\BCBL {}\ \BBA {} Rivera, R.%
\end{APACrefauthors}%
\unskip\
\newblock
\APACrefYearMonthDay{2010}{{\APACmonth{09}}}{}.
\newblock
\APACrefbtitle {{Evoluci\'on y distribuci\'on del ingreso agr\'icola en
  Am\'erica Latina: evidencia a partir de cuentas nacionales y encuestas de
  hogares [Evolution and distribution of agricultural income in Latin America:
  Evidence from national accounts and household surveys]}} {{Evoluci\'on y
  distribuci\'on del ingreso agr\'icola en Am\'erica Latina: evidencia a partir
  de cuentas nacionales y encuestas de hogares [Evolution and distribution of
  agricultural income in Latin America: Evidence from national accounts and
  household surveys]}}\ \APACbVolEdTR{}{\BTR{}\ \BNUM\ LC/W.338}.
\newblock
\APACaddressInstitution{}{{Economic Commission for Latin America and the
  Caribbean}}.
\PrintBackRefs{\CurrentBib}

\bibitem [\protect \citeauthoryear {%
Van~Vlasselaer%
, Eliassi-Rad%
, Akoglu%
, Snoeck%
\BCBL {}\ \BBA {} Baesens%
}{%
Van~Vlasselaer%
\ \protect \BOthers {.}}{%
{\protect \APACyear {2017}}%
}]{%
van2017gotcha}
\APACinsertmetastar {%
van2017gotcha}%
\begin{APACrefauthors}%
Van~Vlasselaer, V.%
, Eliassi-Rad, T.%
, Akoglu, L.%
, Snoeck, M.%
\BCBL {}\ \BBA {} Baesens, B.%
\end{APACrefauthors}%
\unskip\
\newblock
\APACrefYearMonthDay{2017}{}{}.
\newblock
{\BBOQ}\APACrefatitle {Gotcha! Network-based fraud detection for social
  security fraud} {Gotcha! network-based fraud detection for social security
  fraud}.{\BBCQ}
\newblock
\APACjournalVolNumPages{Management Science}{63}{9}{3090--3110}.
\PrintBackRefs{\CurrentBib}

\bibitem [\protect \citeauthoryear {%
Zhang%
, Jia%
, Diao%
, Hai%
\BCBL {}\ \BBA {} Li%
}{%
Zhang%
\ \protect \BOthers {.}}{%
{\protect \APACyear {2016}}%
}]{%
zhang2016}
\APACinsertmetastar {%
zhang2016}%
\begin{APACrefauthors}%
Zhang, Y.%
, Jia, H.%
, Diao, Y.%
, Hai, M.%
\BCBL {}\ \BBA {} Li, H.%
\end{APACrefauthors}%
\unskip\
\newblock
\APACrefYearMonthDay{2016}{}{}.
\newblock
{\BBOQ}\APACrefatitle {Research on {{Credit Scoring}} by {{Fusing Social Media
  Information}} in {{Online Peer}}-to-{{Peer Lending}}} {Research on {{Credit
  Scoring}} by {{Fusing Social Media Information}} in {{Online Peer}}-to-{{Peer
  Lending}}}.{\BBCQ}
\newblock
\APACjournalVolNumPages{Procedia Computer Science}{91}{}{168-174}.
\PrintBackRefs{\CurrentBib}

\bibitem [\protect \citeauthoryear {%
Zhou%
, Tu%
, Chen%
\BCBL {}\ \BBA {} Wang%
}{%
Zhou%
\ \protect \BOthers {.}}{%
{\protect \APACyear {2017}}%
}]{%
zhou2017}
\APACinsertmetastar {%
zhou2017}%
\begin{APACrefauthors}%
Zhou, J.%
, Tu, Y.%
, Chen, Y.%
\BCBL {}\ \BBA {} Wang, H.%
\end{APACrefauthors}%
\unskip\
\newblock
\APACrefYearMonthDay{2017}{{\APACmonth{01}}}{}.
\newblock
{\BBOQ}\APACrefatitle {Estimating {{Spatial Autocorrelation With Sampled
  Network Data}}} {Estimating {{Spatial Autocorrelation With Sampled Network
  Data}}}.{\BBCQ}
\newblock
\APACjournalVolNumPages{Journal of Business \& Economic
  Statistics}{35}{1}{130-138}.
\PrintBackRefs{\CurrentBib}

\end{thebibliography}

\end{document}